\documentclass[conference]{IEEEtran}
\usepackage{times}

\usepackage[numbers]{natbib}
\usepackage{multicol}
\usepackage[bookmarks=true]{hyperref}
\usepackage{graphicx}
\usepackage{MnSymbol}
\usepackage{float}
\usepackage{booktabs}
\usepackage{caption}
\usepackage{subfigure}
\usepackage{syntax}
\usepackage{amsmath}
\usepackage{parskip}

\begin{document}

\title{VidereX: A Navigational Application inspired by ants}




%
\author{\authorblockN{Nam Ho Koh\authorrefmark{1},
Doran Amos\authorrefmark{2},
Paul Graham\authorrefmark{3}, 
Andrew Philippides\authorrefmark{3}}

\authorblockA{\authorrefmark{1}Computer Science \& Engineering\\
University of Michigan - Ann Arbor, MI. Email: namhokoh@umich.edu}

\authorblockA{\authorrefmark{2}School of Life Sciences\\
University of Sussex, Brighton. Email: D.P.Amos@sussex.ac.uk}

\authorblockA{\authorrefmark{3}School of Life Sciences\\
University of Sussex, Brighton. Email: P.R.Graham@sussex.ac.uk}

\authorblockA{\authorrefmark{4}School of Engineering and Informatics\\
University of Sussex, Brighton. Email: andrewop@sussex.ac.uk}}

\maketitle
\begin{abstract}
Navigation is a crucial element in any person's life, whether for work, education, social living or any other miscellaneous reason; naturally, the importance of it is universally recognised and valued. One of the critical components of navigation is vision, which facilitates movement from one place to another. Navigating unfamiliar settings, especially for the blind or visually impaired, can pose significant challenges, impacting their independence and quality of life \cite{chanana2017assistive}. Current assistive travel solutions have shortcomings, including GPS limitations and a demand for an efficient, user-friendly, and portable model \cite{chanana2017assistive, helal2008engineering}. Addressing these concerns, this paper presents VidereX \cite{VidereX}, a smartphone-based solution using an ant-inspired navigation algorithm. Emulating ants' ability to learn a route between nest and feeding grounds after a single traversal, VidereX enables users to rapidly acquire navigational data using a one/few-shot learning strategy. A key component of VidereX is its emphasis on active user engagement. Like ants with a scanning behaviour to actively investigate their environment, users wield the camera, actively exploring the visual landscape. Far from the passive reception of data, this process constitutes a dynamic exploration, echoing nature's navigational mechanisms.

\end{abstract}

\IEEEpeerreviewmaketitle

\section{Introduction}
Many visually impaired and blind suffer from being excluded from activities that strongly depend on navigation and mobility. Consequently, they are discouraged from further participation in social interactions, which may also result in isolation and depression \cite{popescu2012explaining}. In 2020, the World Health Organization estimated that the total number of visually impaired people stood at 285 million worldwide, of which 29 million are blind, and 246 million have low vision \cite{WHO}. For a sighted person to navigate an environment, one can easily acquire enough visual cues consisting of building names, street signs and distinct landmarks, which is crucial in guiding them to their destination. On the other hand, the blind and visually impaired are at a significant disadvantage as they do not have direct access to the same type of information.

According to Brambring’s locomotion model of the blind, perception, landmarks, and orientation are critical to their environmental navigation \cite{hersh2010assistive}. Building on this model, this paper demonstrates that visual cues can be an effective navigational tool when embodied in the VidereX system, which instantiates a variant of our ant-inspired navigation algorithm \cite{baddeley2012model} for use with a standard mobile phone. In particular, VidereX allows for rapid one/few-shot learning of routes because users actively interact with their environment, similar to how the ants the algorithm is based on scan to explore their surroundings actively \cite{wystrach2014visual}. Here we present the VidereX system and show that it is possible for users, even with limited visual input, can rapidly learn and navigate their surroundings through active engagement and exploration, leveraging visual cues for efficient and independent navigation.

\noindent \textbf{Background: visual guidance aids.} Navigating their environment, the blind and visually impaired often resort to traditional aids such as the long cane, guide dogs, and sighted guides to overcome mobility challenges. However, these methods have their shortcomings. Guide dogs, while helpful, are not universally adopted and can impose considerable costs due to potential mismatching issues \cite{lloyd2016investigation}. The long cane offers valuable ground-level information but can lead to awkward reaching behaviours \cite{marston2003hidden}, cause pedestrian tripping in crowded spaces, and offer limited protection above the waist. Furthermore, navigating with impaired sight increases the risk of head-level and fall accidents, often requiring medical treatment \cite{manduchi2011mobility}, consequently reducing confidence in mobility. As for sighted guides, they provide reliable assistance but require significant trust and entail additional costs.
Adding to these challenges is the inherent inadequacy of GPS technology in indoor navigation. GPS signals struggle to penetrate buildings, rendering them less effective indoors. This lack of accurate positioning can pose considerable difficulties for the visually impaired when navigating indoor spaces, further compounding their mobility challenges. Thus, there is a pressing need for more effective and reliable navigation aids for the blind and visually impaired.

Extensive research has been conducted in assistive technology and electronic travel aids \cite{marston2003hidden,manduchi2011mobility,bhowmick2017insight,ran2004drishti} for the visually impaired and blind. This may be attributed to the scope of research extending from the physiological factors pertaining to vision loss to human factors associated with mobility, access to information and various technological developmental aspects in the forms of navigation and way-finding. Due to this expansive research space, authors in \cite{bhowmick2017insight} state that it is a challenge to capture the true essence of this field into one single concept. However, despite the progress in the domain, many of the navigation systems proposed for the blind and visually impaired are brought with various limitations, and only a mere few can truly deliver a dynamic, seamless and robust solution \cite{ran2004drishti}. From analyzing the different solutions present in addressing navigation challenges for the visually impaired and blind, it is clear that many tools often require the user to carry additional equipment and hardware, which might not be as comfortable nor inexpensive \cite{helal2008engineering}. Furthermore, the emergence of several GPS-based solutions may provide specific degrees of accuracy and reliability for blind and visually impaired users' daily navigational tasks. However, due to the inherent flaws of GPS, an alternative solution might be needed.

\noindent \textbf{Background: Familiarity-based navigation.} In this paper, we introduce a navigation model inspired by the behaviours of ants. In their natural environment, ants set out, return via path integration, and on this first route, traversal can acquire the navigational knowledge to repeat this journey using only visual information \cite{mangan2012spontaneous}. This ability is primarily thought to be due to their active strategy of using views as a "visual compass". That is, the ants scan their surroundings, compare what they see to the views they remembered during the first PI-mediated route traversal, and then move in the most familiar direction (i.e. most similar when compared to their memories). This active scanning approach forms the basis for our navigation algorithm. 

\noindent \textbf{Background: Overview.} The challenge we address is translating this ant-inspired strategy to a handheld camera. Unlike ants' panoramic perception, a camera offers a limited field of view, necessitating a sweeping motion to capture the surroundings. We developed a mobile application and algorithm replicating this ant-like active scanning and one-shot learning. With VidereX, a user walks a path, and the visual data is captured and uploaded. Subsequently, a second user actively navigates by continually sampling the visual environment. This paper presents the application and shows proof in principle that this model can be effectively implemented in a mobile application. Specifically, we demonstrate the feasibility and potential of active, visual navigation by showing that a user can actively regain the correct heading by a sweep through the environment.

\section{Methods}

\noindent \textbf{Methods: Familiarity-based navigation.} As detailed in \cite{baddeley2012model}, ant-inspired familiarity-based navigation is a two-phase algorithm encompassing training and execution stages. In the training phase, the agent traverses the route and captures panoramic views to create a comprehensive route memory. In the execution phase, these captured views are used as "visual compasses", where the agent compares their current view to the memories and moves in the most familiar direction by seeing which direction is most similar to its stored memories. The agent identifies the most familiar direction by rotating in place, comparing each rotated version of the current view with its stored snapshots. This process replicates ants' saccadic head movements to find familiar views and decide their next move, allowing them to retrace their path reliably \cite{baddeley2012model,philippides2011might}.

\noindent \textbf{Methods: VidereX algorithm.} The algorithm used in VidereX is a variant of the \textit{Perfect Memory} algorithm \cite{knight2019insect}. In our version, the user performs the training sequence by stepping through the route while recording it on their mobile device. Subsequently, the image frames will be extracted and stored as the main route. When returning to the same route, the user will scan the path with their camera, where the images in memory will be compared against the current frames being passed in real-time.
Moreover, the difference between each frame in memory and the real-time frame being viewed by the user is denoted by an auditory tone, where a higher tone signifies the best match. Therefore, by listening to the tone alone, the user could head towards the correct heading along the trained route. This interaction is represented in figure \ref{fig:outside}. To ascertain the similarity between the current view and each of the training views, we follow Zeil et al. \cite{zeil2003catchment} and use an image difference function (IDF) which gives the difference between a remembered training view or `snapshot' and the current world view (X, Y) respectively \cite{philippides2012can}.

\begin{equation}
\operatorname{IDF}(X, Y)=\frac{1}{P} \sqrt{\sum_{i} \sum_{j}(X(i, j)-Y(i, j))^{2}}
\end{equation}

\noindent where X(i,j) represents the pixel in the ith row and the jth column of image X. This simple RMS measure of pixel difference has been shown to be a reliable cue for visual homing \cite{zeil2003catchment,philippides2011might}. However, \cite{philippides2012can} outlined potential limitations present due to the variation in illumination and environmental changes relating to the time of day and variability of vegetation. 

\begin{figure}[H]
    \centering
    \includegraphics[width=0.4\textwidth]{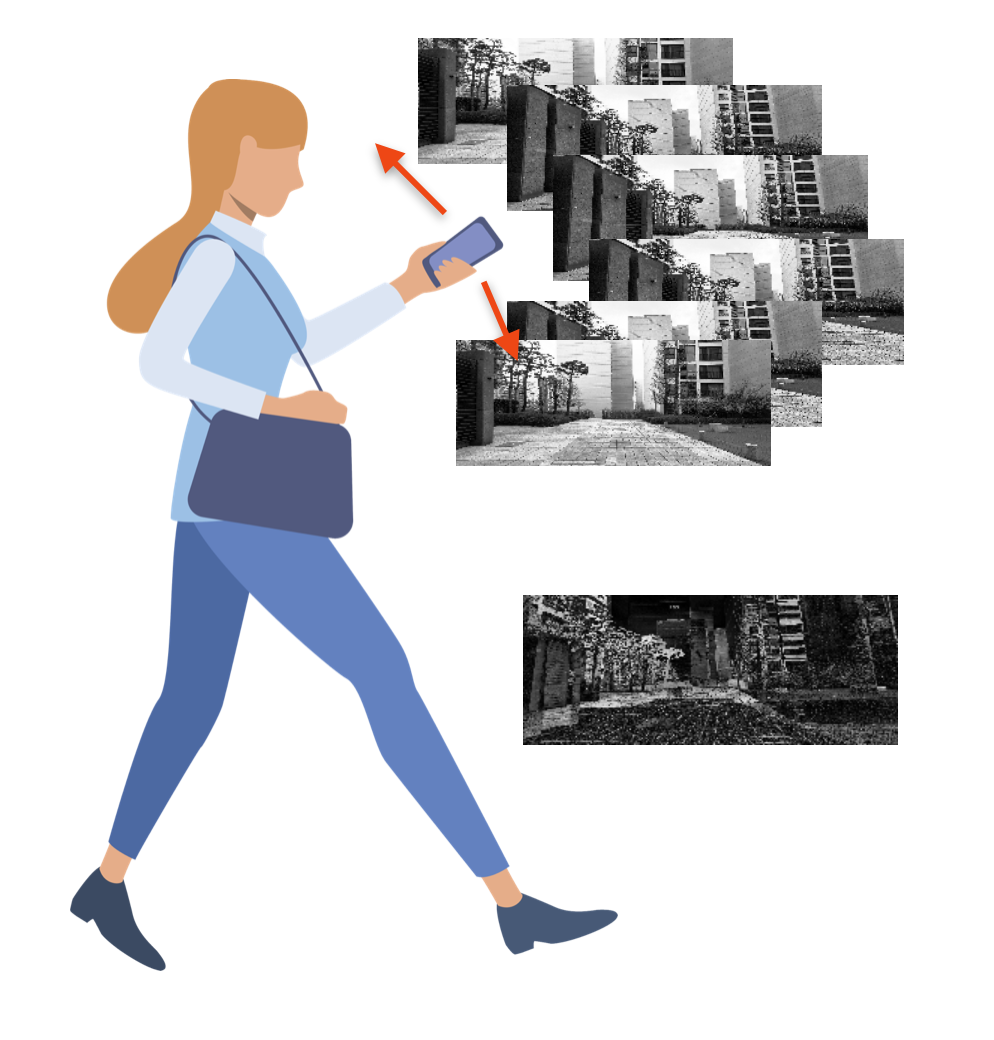}
    \caption{Example of training sequence performed on device. The user is scanning the environment to determine the correct heading from the learnt views from training.}
    \label{fig:training}
\end{figure}

\noindent \textbf{Methods: VidereX Application.}
The final prototype of the application \textit{(VidereX)} contains a rich set of features to augment the BVI navigational experience. In essence, the application will allow the user to perform a single matching action with a single image taken as the reference point, record an entire route via a video, store it on-device and via the cloud using Firebase and follow any route from the start. In addition, the application carries a robust on-device NLU (Natural Language Utterance) system, which facilitates hands-free interactions for increased accessibility user experience.

\noindent \textbf{Methods: Platform and tools.} A Samsung GalaxyS10+ and GalaxyS21 Ultra were during the development and testing cycle. The OpenCV4Android (C++ Native 3.0) was used for all image processing, The ASR (Automatic Speech Recognition) was handled with TensorFlow Lite and the TTS (text-to-speech) feature used Android's native NLU model. 

\begin{figure}[H]
    \centering
    \includegraphics[width=0.46\textwidth]{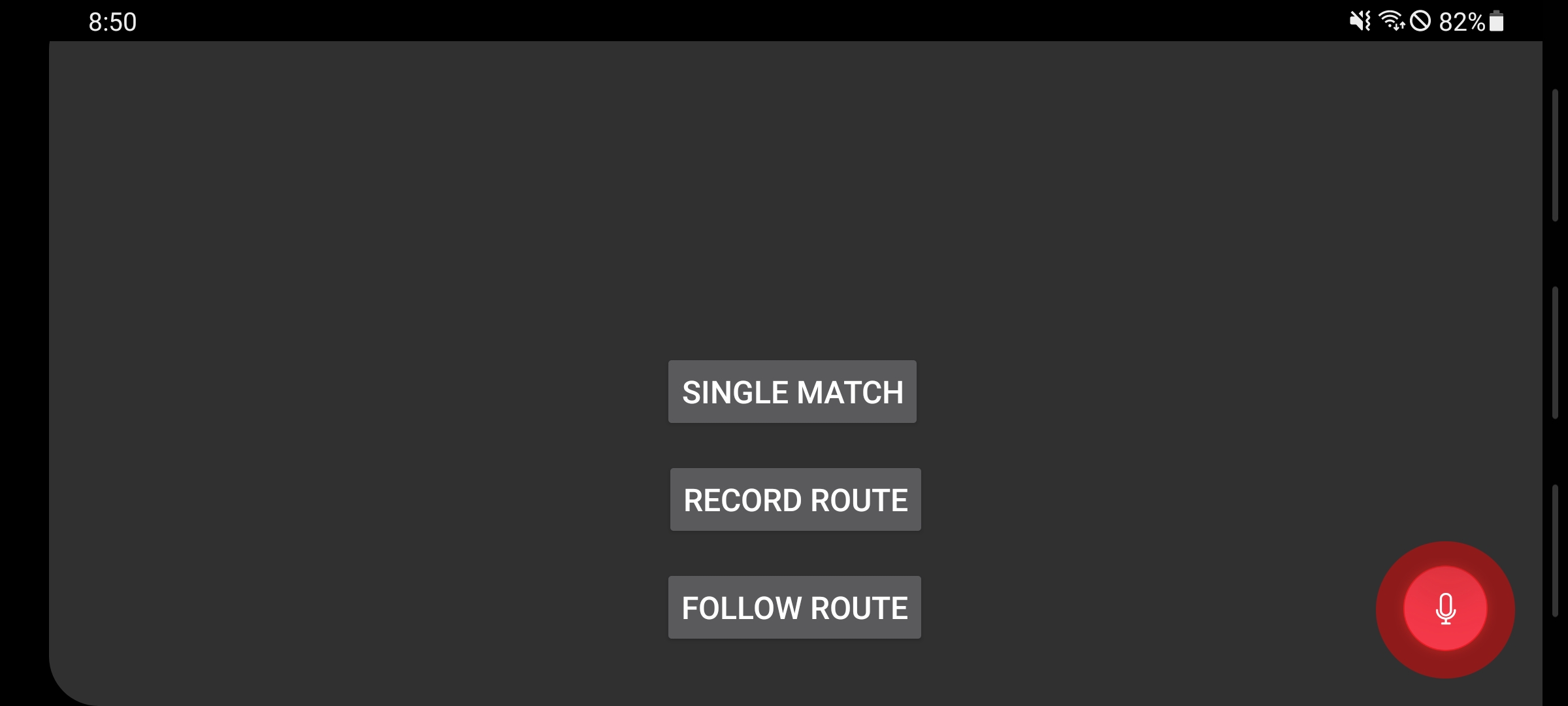}
    \caption{MainMenuActivity screenshot of VidereX. The user is able to execute single match, record route or follow route activities. The menu also holds a speech execution button which will allow the user to perform all actions through natural language commands.}
    \label{fig:route}
\end{figure}

\noindent \textbf{Methods: Single Matching.}
The single match activity allows the user to take a single image and utilize it for visual matching against the environment. The application informs the user via auditory and tactile cues that a match has occurred. The former is set through a tone generator library in which the maximum and minimum difference values from the calculation are normalized to the range of the tone generation frequency. In this case, a low difference value denoting a good match will be met with a high frequency, and in contrast, a high difference value will return a low frequency. The latter takes advantage of the Android vibration library, which triggers when a good match has occurred. The implementation of both signals seemed appropriate, as recommended in \cite{giudice2008blind}.

\noindent \textbf{Methods: Record Route.}
The record route activity allows the user to step and record through an entire route via video recording. Once the video has been recorded, the user can name the route as demonstrated below: 

Once the user has done so, they can extract the frames from the video, which will be used for visual guidance; these frames are stored locally; however, in future implementations, the images will be stored on a cloud instance to minimize storage and increase accessibility. Moreover, once the user has been notified that the frames have been extracted and stored, they can either follow the route directly or store the route for later access. 

\noindent \textbf{Methods: Follow Route.}
The follow route activity allows users to select a stored route on the cloud for navigation. Once a specific route has been chosen via voice or touch, the application will call a data retrieval method to the selected route, as shown in Figure 6.4. Furthermore, if the route exists, the application will pass the respective image paths to the next activity for the images to be used for navigation. Moreover, the route data is collectively stored on Firebase via Firestore \cite{GoogleFire}, a scalable database aimed at mobile and web applications. Its ability to perform atomic batch operations offers a lightweight solution ideal for a smartphone application that relies on low latency and high performance. 

\begin{figure}[H]
    \centering
    \includegraphics[width=0.46\textwidth]{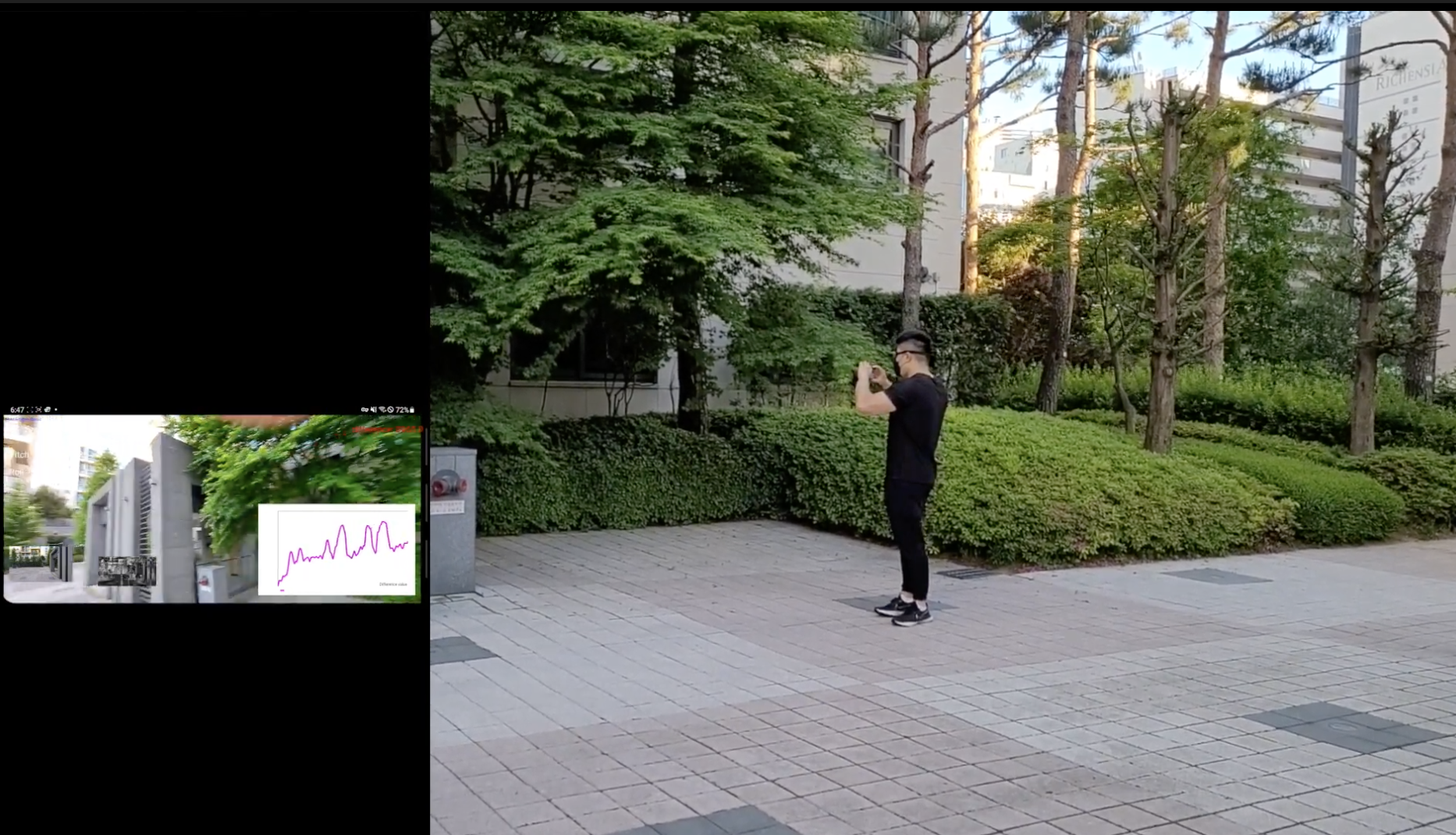}
    \caption{Outside route demonstration.}
    \label{fig:outside}
\end{figure}

\noindent \textbf{Methods: Dataset.}
In order to test the algorithm, we collected 20,418 image samples in varying locations by taking a 180° sweep at respective positions along each route, as shown in Figure \ref{fig:enhanced} below. In addition, 4 unique test positions were taken 7cm apart (approximately one step) with the camera facing directly forwards in parallel.
\begin{figure}[H]
    \centering
    \includegraphics[width=0.4\textwidth]{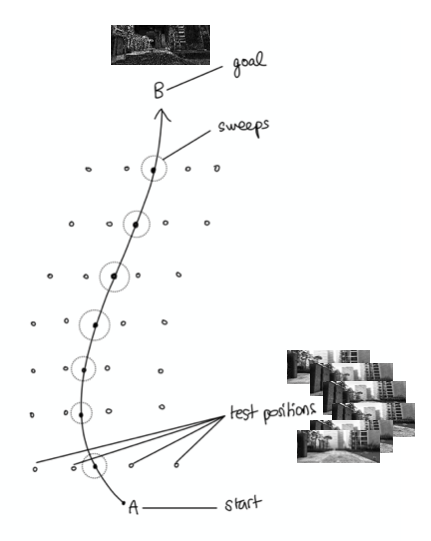}
    \caption{Enhanced data collection route. \textbf{A} denotes the starting position,  \textbf{B} represents the goal position. The dotted circles represent the rotational sweeps and the unmarked circles left-right of the sweeps represent the individual test positions taken.}
    \label{fig:enhanced}
\end{figure}
\vspace{-0.2cm}
The purpose of this data collection method was to augment the data by capturing different angular positions and distances as well as allowing the algorithm to be tested across different environments and scenery.

\section{Results}
\begin{figure}[H]
    \centering
    \includegraphics[width=0.45\textwidth]{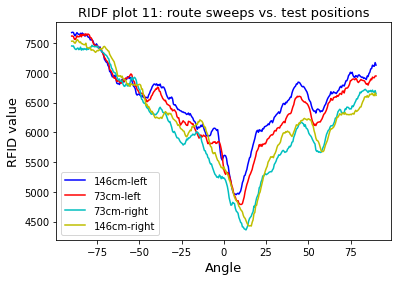}
    \caption{RIDF plot taken from route rotational sweep against the test position. General heading can be derived shown by the widened minima of the functions with similar patterns. This suggests that each instance captures synonymous visual cues and information from the environment at each position.}
    \label{fig:ridf}
\end{figure}

\noindent \textbf{Results: Testing environment.} Our initial test of the algorithm sought to assess its capability in a real-world setting. We were particularly interested in understanding whether a user, actively searching, could successfully identify the correct path with the aid of our algorithm. Furthermore, to gauge the robustness and accuracy of our navigation solution, we tested positions at different distances from the training path. The results of these initial trials are shown in fig \ref{fig:ridf}. Here we see that a general heading can be derived, as shown by the widened minima of the RIDFs. This suggests that each instance captures synonymous visual cues and information from the environment at each position.

\begin{figure}[H]
    \centering
    \includegraphics[width=0.45\textwidth]{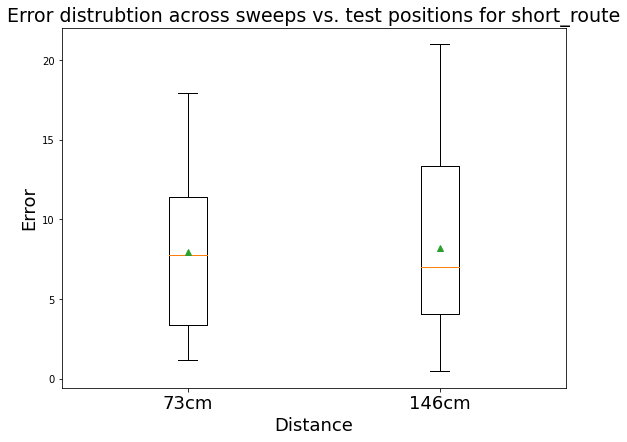}
    \caption{Boxplots of error distribution across rotational sweeps and test positions along the different routes.}
    \label{fig:bp1}
\end{figure}

\vspace{-0.8cm}
We found that the angular error slightly increased as the user deviated from the centre of the sweep. However, the deviations remained within acceptable ranges, thus demonstrating the potential for practical yet reliable navigation. The Rotational Image Difference Function (RIDF) depicted in figure \ref{fig:ridf} confirms the presence of the necessary visual information that permits the user to find the true heading despite the sweep being performed manually, as we can see the clear minimum near the true heading. In addition, these results remain consistent across different environments as the error distribution, visualised in the box plots of figure \ref{fig:bp1}, shows the algorithm's effectiveness and robustness at different distances from the route. 
It is crucial to mention that although the algorithm is extremely promising in its wide applicability in navigation, continuous rigorous testing is needed to prove the algorithms worth.

\section{Conclusion} 
\label{sec:conclusion}
Support and care for the blind and visually impaired have been progressively improving worldwide, mainly due to technological advancements and the development of assistive navigational solutions. While these improvements are noteworthy, our vision, often taken for granted due to its constant presence, is an invaluable tool that is profoundly missed when absent.
This paper shows that studying the visual behaviour of ants in their natural environment can provide a solution in a very different context by translating their remarkable navigation strategies into a smartphone application for the blind and visually impaired.

Exploring this algorithm's implementation deepened the authors' understanding of assistive technology solutions and shed light on the various user constraints in this underrepresented global population. Navigation, whether for sighted or non-sighted individuals, remains a fundamental principle that empowers and liberates individuals. However, as we look to the future, there are improvements to make to VidereX to make it more robust. Collecting more data and conducting detailed testing is the first step to allow us to refine the algorithm, enhancing its precision and reliability. Additionally, incorporating additional modalities, such as an inertial measurement unit (IMU), could enhance the system's navigational capabilities by restricting matches to only those where images have similar attitudes. 

In summary, while our insect-inspired navigation algorithm implemented in a smartphone application is functional and promising, it represents a starting point. The continued exploration, development, and improvement of such assistive technology solutions could be crucial to promoting the independence and quality of life of the blind and visually impaired.

\section*{Acknowledgments}
We thank the EPSRC (Engineering and Physical Sciences Research Council) for funding Andy Philippides and the CCNR (Centre for Computational Neuroscience and Robotics) for the support.


\bibliographystyle{plainnat}
\bibliography{references}

\end{document}